\documentstyle[aas2pp4]{article}
\begin{document}
\lefthead{Kawai {\it et al.}}
\righthead{PCI-HIB}
\title{The PCI Interface for GRAPE Systems: PCI-HIB}

\author{
Atsushi Kawai$^{1}$,
Toshiyuki Fukushige$^{1}$,
Makoto Taiji$^{2}$,\\
Junichiro Makino$^{1}$ and
Daiichiro Sugimoto$^{1}$
}
\affil{
$^1${
	Department of General Systems Studies,
	College of Arts and Sciences,
	University of Tokyo,
}\\
{
	3-8-1 Komaba,
	Meguro-ku,
	Tokyo 153
}\\
$^2${Institute of Statistical Mathematics,} \\
{
4-6-7 Azabu,
Minato-ku,
Tokyo 106
}
}
\authoremail{kawai@chianti.c.u-tokyo.ac.jp}

\begin{abstract}
We developed a PCI interface for GRAPE systems.
GRAPE(GRAvity piPE) is a special-purpose computer for gravitational
$N$-body simulations. A GRAPE system consists of GRAPE processor
boards and a host computer. GRAPE processors perform the
calculation of gravitational forces between particles. The host
computer performs the rest of calculations.  The newest of GRAPE
machines, the GRAPE-4, achieved the peak performance of 1.08 Tflops.
The GRAPE-4 system uses TURBOChannel for the interface to the host,
which limits the selection of the host computer. The TURBOChannel bus
is not supported by any of recent workstations.
We developed a new host interface board which adopts the PCI bus
instead of the TURBOChannel.  PCI is an I/O bus standard developed by
Intel. It has fairly high peak transfer speed, and is available on
wide range of computers, from PCs to supercomputers. Thus, the new
interface allows us to connect GRAPE-4 to a wide variety of host
computers.
In test runs with a Barnes-Hut treecode, we found that the performance
of new system with PCI interface is 40\% better than that of the
original system.
\end{abstract}
\keywords{
	Clusters: globular --- Numerical methods --- Stars: stellar dynamics
}

\section{Introduction}

Many astronomical objects can be well approximated by gravitational
$N$-body systems. To understand the behavior of gravitational $N$-body
systems is one of the most important problems in theoretical
astrophysics.

Numerical simulation is widely used to study the behavior of $N$-body
systems, because in many cases the analytic approach or more idealized 
method are not sufficient. However, in many fields the limitation in
the resolution and accuracy makes it difficult to obtain meaningful
results from $N$-body simulations. These limitations come mainly from
limited number of particles.

In $N$-body simulations, the calculation speed of the computer is the
primary cause to limit the number of particles, since a naive
algorithm requires $O(N^2)$ calculation cost per timestep, where $N$
is the number of particles. The calculation cost is $O(N^2)$ because
gravity is a long-range interaction. Each particle interacts with all
other particles.

In some cases, approximate algorithms such as the P$^3$M algorithm
and the Barnes-Hut treecode (Barnes and Hut 1986) can be used,
resulting in the reduction of calculation cost from $O(N^2)$ to $O(N \log
N)$. Even with these schemes, the cost of the gravitational
interaction is the dominant part of the total calculation cost. There
are a number of studies to improve the efficiency of these schemes, in
particular on parallel computers.

We have been exploring an alternative approach, which is to develop a
special-purpose hardware for the force calculation. The calculation of
the force between two particles is simple enough to be put on a
hardwired pipeline (Sugimoto et al. 1990). This hardware works in
cooperation with a general-purpose computer (host computer), which
does everything except for the force calculation. In the case of the
direct force calculation, calculation cost of the force is $O(N^2)$
and that of the rest is $O(N)$. Therefore the requirement for the
speed of the calculation on the host and the data transfer between the
host and the special-purpose hardware is rather modest. In addition,
fast, approximate schemes such as the P$^3$M and Barnes-Hut treecode
can be further accelerated with the hardware for the particle-particle
force (Makino 1991a, Brieu et al. 1994), though in this case the
requirement for the performance of the host becomes somewhat higher
simply because the total calculation cost is smaller than that for the
direct summation.

The GRAPE-4 system (Taiji et al. 1996, Makino et al. 1997) is our
newest hardware with the theoretical peak speed of 1.08 Tflops. The
measured best speed so far was 523 Gflops, for the simulation of black 
hole binaries in the center of an elliptical galaxy (Makino and Taiji 1995).

In this paper, we describe the enhancement we added to GRAPE-4 to
further improve its performance on real problems. In many simulations,
the performance of GRAPE-4 was limited by the speed of the host
computer. This is essentially because GRAPE-4 is so fast,
but partly because we could not use the fastest host computer
available. The hardware interface to the host is designed around
TURBOChannel, the I/O bus specification developed by DEC. At the time
of the development of GRAPE-4, it was a reasonable choice, but in 1994 
DEC dropped the product line with TURBOChannel. 

In order to improve the performance of the host computer, we designed
a new host interface for GRAPE-4. We adopted the PCI interface as the
host interface.  PCI (Peripheral Component Interconnect, PCI Special
Interest Group 1993) is an I/O bus standard developed by Intel. It is
currently the most widely used I/O bus for PCs with Intel x86
processors, and a number of computer manufacturers of all kinds of
products, including vector supercomputers, massively parallel
computers, SMP servers and workstations, have shifted from proprietary
I/O bus to this PCI bus.  Unlike the TURBOChannel, PCI will be around
for next 5-10 years, if we can judge its lifetime from that of its
predecessor, the ISA bus.

As of 1996, virtually all manufacturers of PC are providing PCI, and
all major workstation vendors either are shipping PCI products (DEC,
HP, SGI, IBM) or have announced products (SUN). In addition, Apple has
also shifted from NuBus to PCI in 1994. Recently announced NEC SX-4A
vector supercomputer also supports PCI as I/O interface.

There are several reasons why the PCI bus is now supported on almost
all computers. The first one is the production cost. PCI is first
adopted by PC manufacturers, which are now producing more than 90\% of 
the total number of computers. Thus, the peripherals such as display
cards, SCSI interface cards and network interface cards with PCI
interface are produced in the quantity more than 10 times larger than
the total of all other kinds of peripherals, and therefore the
production cost of the PCI-based interface is much lower. By adapting
the PCI bus, any computer can use the peripherals  designed for PCs,
thereby greatly reducing the total development cost.

The second reason is that the performance of the PCI interface is
quite high. Even with the low-end specification, it offers the maximum
data transfer rate of 133 MB/s, which is more than enough for almost
any storage or communication device. A high-performance hard disk unit
would offer the data transfer rate of 10-20 MB/s.  Fast Ethernet and
ATM network also offer the data transfer rate of 10-15 MB/s. Thus, the
speed of 133 MB/s is more than enough. Even so, high-end PCs and
workstations now support 266MB/s, and the data transfer rate of 533
MB/s might be implemented soon.  This transfer speed is faster than
those of other proprietary I/O buses.

The wide variety of computers which support PCI and the high
performance of the PCI makes it an ideal choice as the I/O interface
for GRAPE hardwares, including the ones we will develop in future. On
the other hand, to develop a PCI interface is a challenging task,
partly because of its high performance and partly because of its
complexity. 

In this paper, we describe the implementation and the performance of
PCI interface hardware we developed for GRAPE-4.  In section 2, we
briefly describe the GRAPE-4 system.  In section 3, we describe the
PCI interface for GRAPE-4. We call it PHIB (PCI host interface board).
In section 4, we present the measured performance of PHIB and GRAPE-4
with PHIB. Section 5 is devoted for discussions.

\section{GRAPE-4 System}

In this section, we describe the GRAPE-4 system. In section 2.1, we
show the hardware structure of the system. In section 2.2, we describe
how the actual $N$-body simulation is performed with this system.  In
section 2.3, the function of HIB is described.

\subsection{The Hardware Structure}

\begin{figure}
\plotone{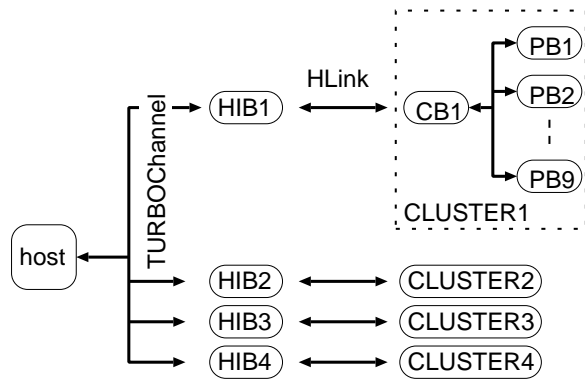}
\caption
{
The GRAPE-4 System
}
\end{figure}

Figure 1 shows the structure of the GRAPE-4 system. It consists of a
host computer and multiple clusters. Each cluster consists of a HIB
(Host Interface Board), a CB (Control Board), and PBs (Processor
Boards). DEC Alpha AXP 3000/x00 machines are used as the host
computer, and HIBs are attached to the DEC TURBOChannel bus. For the
connection between HIB and CB, we used a data transfer protocol which
we call Hlink (HARP-link). It is a synchronous protocol with 32-bit
data width. CB and PBs are connected by the custom bus with 96-bit
data width and synchronous protocol with fixed latency, which we call
HBus (HARP-Bus).

PBs calculate gravitational forces with parallel pipelines. A PB
contains 47 HARP-Chips each of which has two virtual pipelines to
calculate gravitational forces and operates on 32 MHz clock cycle. One
PB contains 94 virtual gravitational pipelines which operate on 16MHz
clock cycle.  CBs sum up the forces calculated on PBs and return the
results to the host.  HIBs translate the TURBOChannel protocol to
Hlink protocol and vice versa.

\subsection{$N$-body Simulation with the GRAPE-4 System}

The simplest $N$-body simulation with the GRAPE-4 system proceeds in
following order:

\renewcommand{\labelenumi}{\arabic{enumi})}
\begin{enumerate}
\item{
The host sends the information of $N$ particles to the memory on PBs
where $N$ is the number of the particles in the system.  We call this
operation ``$j$-particle write operation''. For brevity, we call the
particles of which data are stored in the memory of PB as
``$j$-particles''.
}

\item{
The host sends the information of $n_{\rm vp}$ particles to the
pipelines on PBs. Here $n_{\rm vp}$ is the number of virtual pipelines
on one PB, which tipically equals to 94. We call this operation ``$i$-
particle write operation''. We call the particles to be stored on the
registers as ``$i$-particles''.
}

\item{
The pipelines on PBs calculate the forces for $n_{\rm vp}$ $i$-particles.
}

\item{
The host receives $n_{\rm vp}$ calculation results from pipelines on PBs.
We call this operation ``result read operation''.
}

\item{
The procedures 2) to 4) are repeated until the calculation results for all
$i$-particles are received.
}

\item{
The host updates the positions of all particles in the system.
}

\end{enumerate}

\subsection{The Funciton of HIB}

In the following, we describe the function of HIB in some details,
since its functionality is replicated in the new PCI version of HIB
which will be described in the next section.

HIB communicates with the host in four different ways, namely
programmed (PIO) write, PIO read, DMA read and DMA write. Here,
programmed read/write means that the host computer initiates the
transaction, while DMA means HIB initiates the
transaction. PIO write is used by the host to send single-word
commands to CB or PB, and to set the HIB registers for DMA
operation. PIO read is used only to read the status register of
HIB. DMA read is used for data transfer to CB/PB from the host, and
DMA write is used for data transfer from CB/PB to the host.

This rather complex implementation is quite different from that of the
I/O interface of older GRAPE systems, which used PIO read/write
operations on VME interface to perform all data transfer.  We used DMA
to achieve higher performance. In the case of the TURBOChannel bus on
DEC Alpha AXP systems, the data transfer speed close to the
theoretical maximum of 100 MB/s can be achieved with DMA, while the
throughput of PIO read is around 10 MB/s.  However, the necessary
bandwidth for GRAPE-4 is close to 100 MB/s, and faster data transfer
would further improve performance. Thus, the limited bandwidth of PIO
operation on TURBOChannel was not satisfactory. The hardware to
implement DMA operation on a TURBOChannel card was relatively simple,
since TURBOChannel is an exceptionally simple I/O bus. For example,
the host CPU or any other device cannot interrupt a DMA operation, so
there is no need to handle interrupt and restart of a DMA operation.
There is no need for arbitration since each card slot has its
dedicated signal lines for DMA request/grant.  Of course, this rather
simple protocol comes with several limitations. A single DMA operation
must not transfer more than 128 words, and it should not go across the
2 KB address boundary.

We decided to design the interface between HIB and CB to be
independent of the TURBOChannel protocol. The HIB hardware takes care
of all the limitations of the TURBOChannel protocol.  To hide the
limitations of the TURBOChannel bus, HIB implements a bidirectional
data buffer, which is large enough to hold all data to be moved
between the host and CB/PB by single request.

The data transfer in Hlink has three different modes. The first one is 
single-word handshaked transfer from HIB to CB, which is used to send
commands. The second one is the burst transfer from HIB to CB, and the 
last one is the burst transfer from CB to HIB. The bursts can be
interrupted by the sender, but the receiver must always accept the
data. The way the transfer takes place and the length of the bursts
are determined by the command sent from HIB to CB.

\section{The PCI-HIB}

In this section, we describe the PCI-HIB (PHIB for short) which
replaces the TURBOChannel version of the HIB (THIB for short)
described in the previous section. First we briefly overview the PCI
bus, and compare it with other widely used standard I/O buses such as
ISA, VME, SBus and the TURBOChannel. Then we describe the design
principle and the hardware of the PHIB in detail.

\subsection{The PCI Standard}

The PCI interface was proposed by Intel in 1991 and standardized in
1992 as PCI version 1.0. Version 2.0 was defined in 1993 by PCI
Special Interest Group(PCI Special Interest Group 1993) and the
current standard is version 2.1. The PCI interface is a synchronous,
high-performance I/O bus which has many advantages over other I/O
buses. These advantages include
\begin{description}
\item{$\cdot$} Very high throughput, up to 533 MB/s in the case of
66MHz, 64-bit transfer.

\item{$\cdot$} Low manufacturing cost because of the use of the
card-edge connector.

\item{$\cdot$} Low power consumption because of the elimination of
passive terminators. 

\item{$\cdot$} Form factor similar to that of the ISA bus, which
allows co-existence of PCI and ISA cards in one box.

\item{$\cdot$} Sophisticated transaction protocols such as the
support for the cache coherency.

\item{$\cdot$} Support for automatic configuration which eliminates
the need to set any jumper switches on board. 

\end{description}
Partly because of these technical advantages and partly because of the
fact that PCI is proposed by Intel, which is the largest manufacturer
of microprocessors, PCI has quickly become a widely accepted standard.
Table 1 summarizes the comparison between PCI and several
other I/O specifications.

%
%

%
\begin{deluxetable}{lcrrrcrr}
\tablecaption{Comparison between PCI and other I/O specifications.}
\tablehead{
\colhead{{bus name}} &
\colhead{{synchronous/}} &
\colhead{{clock}} &
\colhead{{bus}} &
\colhead{{transfer}} &
\colhead{{arbitration}} &
\colhead{{pin}} &
\colhead{{physical}}\nl
\colhead{} &
\colhead{{asynchronous}} &
\colhead{{cycle}} &
\colhead{{width}} &
\colhead{{rate}} &
\colhead{{type}} &
\colhead{{count}} &
\colhead{{dimemsion}}\nl
\colhead{} &
\colhead{} &
\colhead{{(MHz)}} &
\colhead{{(bit)}} &
\colhead{{(MB/s)}} &
\colhead{} &
\colhead{} &
\colhead{{(mm)}}
}

\startdata

ISA & synchronous & 8 & 16 & 8 & centralized & 88 & 107 $\times$ 334 \nl
VME & asynchronous & 10 & 32 & 40 & centralized & 82 & 100 $\times$ 160\nl
NuBus & synchronous & 10 & 32 & 40 & centralized & 50 & 102 $\times$ 327\nl
TURBOChannel & synchronous & 25 & 32 & 100 & centralized & 44 & 117 $\times$ 144 \nl
SBus & synchronous & 25 & 64 & 200 & centralized & 66 & 84 $\times$ 142\nl
PCI & synchronous & 66 & 64 & 533 & centralized & 102 & 107 $\times$ 175 \nl
Futurebus+ & asynchronous & 100 & 128 & 1600 & centralized/distributed & 206 & 265 $\times$ 297\nl
\enddata
\end{deluxetable}

One problem with the PCI interface is that it is by far more complex
compared to any other bus standards except Futurebus+. The support for
the automatic configuration makes it necessary to implement rather
complex logic for initial configuration process. Thus, even a simple
slave interface board needs to implement rather complex function. A
master interface board is still more complex, because the DMA transfer
of PCI is interruptable.

\subsection{Design Principles of PHIB}

The goal for the design of the PHIB is quite simple. It needs to
implement all the necessary functions available on the THIB. Even so,
there are many considerations.

A rather important technical decision is whether to use the DMA
transfer or not. In the case of the THIB, we had to use DMA since the
performance of PIO read/write was unacceptably low. However, in the
case of the PCI on workstations or PCs, the performance of PIO
operation is not much different from that of the DMA transfer. For
example, DEC Alphastation 5/600 offers the PIO write throughput of 119
MB/s and read throughput of 56 MB/s, according to the published
document from DEC (Zurawski et al. 1995). PIO write performance in
excess of 80 MB/s has been reported on a Pentium-based PC (Ichikawa
and Shimada 1996).

The above numbers are still somewhat lower than what can be achieved
with DMA. However, in most of recent machines, DMA operation needs
additional data moving in the main memory which further degrades the
effective transfer speed. The DMA operation can only move a block of
the data in main memory, which must be physically in the main
memory. This means that the data must be physically moved between the
main memory and the cache. Unfortunately, the data transfer speed
between the main memory and the cache is not much different from the
speed of the I/O bus. Thus, a factor-of-two or more loss in
performance is associated with the DMA operation.

Another problem with the DMA operation is that the software becomes
much more complicated and error-prone.  In order for an interface card
to perform the DMA operation, the card needs to know the physical
address of the memory to access. Since almost all operating systems
now support the virtual memory, the user program knows only the
virtual address of the data and has no knowledge about the physical
memory. Of course the OS kernel keeps track of the mapping between the
virtual address and physical address, but in order to access this
information, we have to write software. Moreover, if a page fault
occurs, the OS might change this mapping, or even put out a virtual
page to the disk, without notifying the user process. If the OS has
the support for real-time operations, it can prevent the specified
memory region from being paged out, thereby preserving the mapping
between the physical and virtual addresses. However, this
functionality is not available in many popular UNIX systems, or rather
awkward to use.

In the case of the PIO operation, the interface card need not know
anything about the page mapping and the user process can simply
read/write the virtual address assigned to the interface card, through
usual load/store instruction, in other words, through assignment
statement in any high level language. The main body of the software is
portable among different platforms (Makino and Funato 1993).

For the present version of PHIB, therefore, we decided to use only the 
PIO operation. Thus, the data transfer which was performed using DMA
in THIB is at present performed using PIO.

Even though we decided not to use the DMA operation, the
implementation of the PCI interface is still rather
complicated. Therefore we decided to use a general-purpose PCI
interface chip, the PCI9060 from PLX technology Inc., as the interface
to the PCI.

The PCI9060 chip integrates both the slave (for PIO) and master (DMA)
functionalities in a single chip. In addition, it has all the
necessary registers and logics for the automatic setup, thus
eliminating any need for designing logics specific to PCI. Its
interface to the local bus is designed so that it can be directly
connected to the CPU bus of the Intel i960 chips and act as both a DMA
master and a PIO slave. In the present PHIB, the PCI9060 chip acts as
a PIO slave to the host computer and as a DMA master to the local bus.

The PCI9060 interface chip can acts as the DMA master for both the
PCI bus and i960 local bus simultaneously. Thus, the PHIB can perform
the DMA operation to the PCI bus without any change in the design.
This possibility allows us to design the PHIB with extra safety:
If the actual performance of PIO is less than satisfactory, we can
always implement the software for DMA to improve the performance.

\subsection{The Hardware Design}

\begin{figure}
\plotone{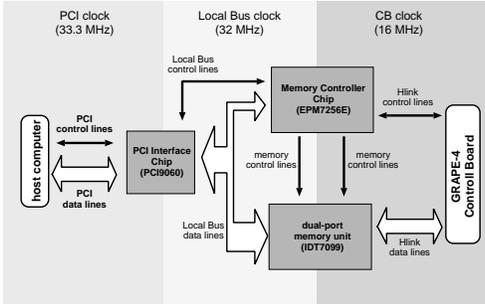}

\caption[hib.eps]
{
The block diagram of the PCI Host Interface Board
}
\end{figure}
%

\begin{figure}
\plotone{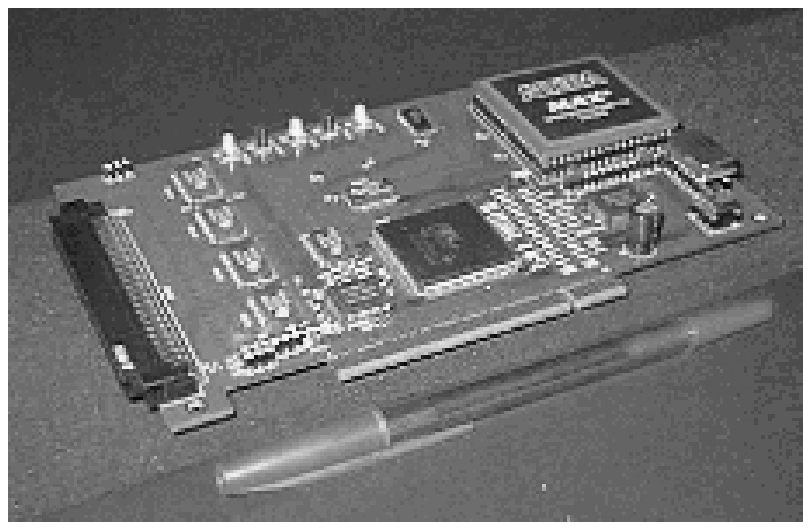}

\plotone{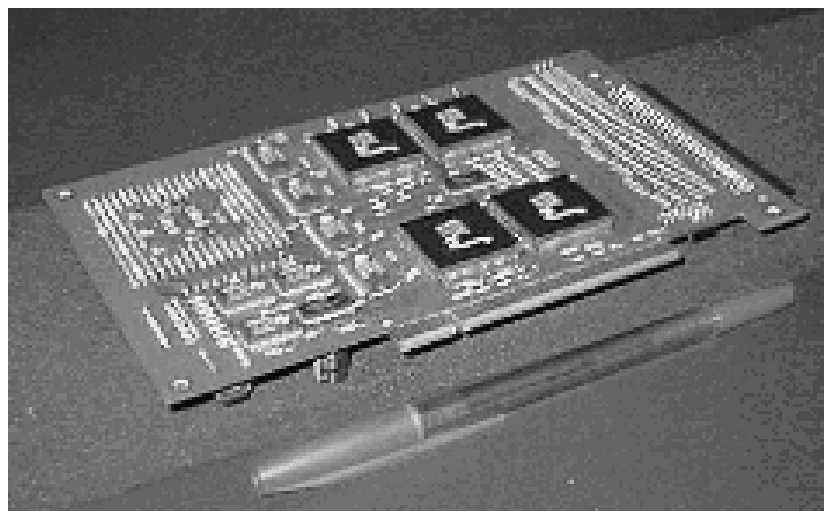}

\caption
{
The PCI Host Interface Board. (a)The parts side. (b) The solder side.
}
\end{figure}

Figure 2 shows the block diagram of PHIB. It consists of the PCI9060
chip, the Memory Controller chip (MC), and the dual-port memory unit.
Figure 3 shows the photograph of the PHIB board.

The data transferred between the host and GRAPE are buffered by the
dual-port memory unit. MC generates the addresses and other control
signals for both ports of the memory, as well as the handshake signals
to PCI9060 chip and the Hlink.

In THIB, we used the FIFO chips for the data buffer, which are much
easier to use than the dual-port memory. We do not have to provide the
address for FIFO, and it can tell whether there are data to be sent or
not. In addition, it is relatively easy to design the system in which
the two ports of the FIFO chip operate on two independent clocks,
since the flag logic implemented in the FIFO chip takes care of
asynchronous two clocks.

In PHIB, however, we were forced to use the dual-port memory.  The
reason is that the PCI interface need to support the read prefetch in
order to allow the host to perform a burst read operation. In the case
of a PCI burst read operation, the master can terminate the burst at
any moment. Thus, the PCI9060 chip need to prefetch the data to
supply, and when the burst is stopped, data already prefetched are
discarded.  Since some of the data read from the buffer will be
discarded, we cannot use the FIFO as the data buffer. To read the same
data more than once from an FIFO device is rather difficult.
Therefore, we used dual-port memory chips. We used four IDT7099 chips
from Integrated Device Technology, each of which has 8-bit width and
4k-word depth.

\begin{figure}
\plotone{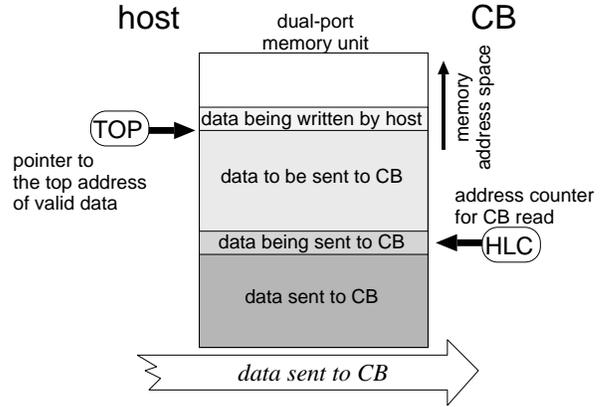}

\caption
{
Data transfer from the host to CB. the HLC(Hlink-port address
counter) and the TOP register in MC are used in this transfer. CB
continues to read from the host computer until HLC catches up with
TOP. The host modifies the TOP register to indicate CB that data to be
transferred is ready.
}
\end{figure}

\begin{figure}
\plotone{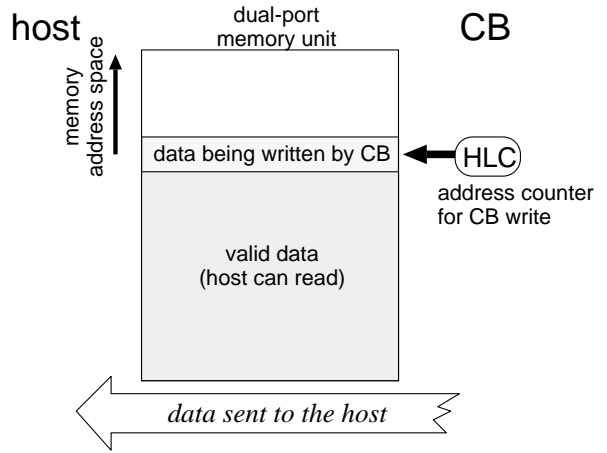}

\caption
{
Data transfer from CB to the host. HLC register is used in this
transfer. HLC points to the address of dual-port memory at which CB
write data.
}
\end{figure}

In the case of the data transfer from the host to CB, the host first
stores the data to the dual-port memory and then sets the register
within MC to be the top address of the region with valid data. MC then
sends out the new data to CB automatically. Figure 4 shows how the
data transfer from host to CB proceeds.

In the case of the data transfer from CB to the host, when PHIB
receives a data, it stores the received data to the dual-port memory
and increments the address counter. The host computer tests if new
data are ready or not by reading this counter, and if there are new
data it reads them in. Figure 5 shows how the data transfer proceeds.

The interface to Hlink operates on the clock signal from CB. The
interface to PCI (part of the PCI9060 chip) operates on the PCI
clock. The local bus of the PCI9060 operates on the clock independent
of either of two external clocks. The speed of these three clocks are
16MHz, 33.3MHz, and 32MHz, respectively. Figure 2 shows which parts of
the PHIB operate using which clock. PCI9060 accept the PCI and Local
Bus clock. Part of MC and one port (Local Bus side) of the dual port
memory operates on the LB clock. The remaining part of MC and the dual
port memory operate on the clock from Hlink. MC unit is implemented in
a single Altera EPM7256E complex PLD chip.

The data width of both PCI and Hlink are 32-bit. Thus the theoretical
maximum of the transfer rate for these buses are 134MB/s and 64MB/s,
respectively.

\section{Performance}

In this section, we describe the performance of PHIB and compare it
with that of THIB on real calculation. In section 4.1, we give the raw
speed of the PCI interface itself, and discuss the dependence on the
host. In section 4.2, we present the data transfer performance of
library functions. In section 4.3 through 4.5, we show the performance
of the direct-summation equal-timestep algorithm, direct-summation
individual-timestep algorithm, and that of Barnes-Hut treecode,
respectively.

\subsection{Raw Data Transfer Speed}

%
%
%
\begin{deluxetable}{lrr}
\tablecaption{PHIB performance}
\tablehead{\colhead{host} & \colhead{Write} & \colhead{Read}}

\startdata

AS500(21164/500MHz) & 68 MB/s & 28 MB/s \nl
PC(P6/200MHz) & 58 MB/s & 10 MB/s \nl
\enddata
\end{deluxetable}

Table 2 shows the raw speed of the data transfer between the host and
PHIB. Here, we simply measure the speed at which the host reads/writes
a continuous region of the dual-port memory on PHIB.  Alpha denotes an
Alphastation 5-500/500 with 500MHz DEC 21164 processor and DEC 21172
PCI chipset. PC is an Intel P6 box from a Japanese company with a
200MHz Pentium Pro processor and Intel 440FX PCI chipset.

For the Alpha box, the performance is satisfactory, if not ideal.  For
the Intel box, however, the performance is rather low, in particular
for the read operation. The difference of the speed of the write
operation between these two hosts is not very large.  The reason for
the rather big difference in the speed of read operation is that the
Alpha box performs the PCI read burst for up to 32 bytes, while the
particular Intel box we used performs only single-word (4 bytes)
reads. The Alpha 21164 CPU can perform the load-merging, which
combines consecutive load requests to the PCI bus to a single burst of
up to 32 bytes. Thus, if the PCI device is fast enough, the data
transfer speed of up to 56MB/s can be achieved. On the other hand, the
PCI interface chipset on the Intel box we used (Intel 440FX) does not
have the capability of load-merging (or at least we could not figure
out how to let that chip do the load merging. Intel documents give
rather little information).  At least one PCI chipset for Intel
Pentium processor seems to perform load-merging (ALI M1448/1449
chipset). We hope similar chipsets to appear for Pentium Pro
processors soon.

The read/write speed of 28MB/s and 68MB/s measured on the Alpha box
are 50\% and 60\% of the maximum values in DEC document, respectively. 
These performance degradation is mainly because of the access latency
of PCI9060 and MC. Taking these latencies into account, the actual
throughput $P_{\rm raw}$ can be approximated as
\begin{equation}
	\label{raw}
	P_{\rm raw} = P_{\rm max}\frac{B}{B+L_{\rm 9060}+L_{\rm mc}+L_{\rm host}}
\end{equation}
where $P_{\rm max} = $ 133MB/s is the maximum data transfer speed of
PCI and $B$ is the length of the PCI burst in clock counts. On the
Alpha box we used, $B$ = 8. The parameters $L_{\rm 9060}$, $L_{\rm
mc}$, and $L_{\rm host}$ are the access latencies of PCI9060, MC, and
the host, in unit of PCI clock. The values for these parapeters are
shown in table 3 both in unit of ns and of PCI clock. Equiation (1)
gives the read/write speed of 30MB/s and 76MB/s. These values agree
well with the measured performance.

%
%
%
\begin{deluxetable}{lrr}
\tablecaption{Access latency of PCI9060, MC, and the host.}
\tablehead{\colhead{Latency source} & \colhead{Write} & \colhead{Read}}

\startdata
PCI9060  & 60 ns (2 clocks) & 390 ns (13 clocks)\nl
MC       & 90 ns (3 clocks) & 120 ns (~\,4 clocks)\nl
host     & 30 ns (1 clocks) & 330 ns (11 clocks)\nl
\enddata
\end{deluxetable}

\subsection{Data Transfer Speed of Library Functions}

In this section we present the data transfer performance of library
functions.


\begin{figure}
\plotone{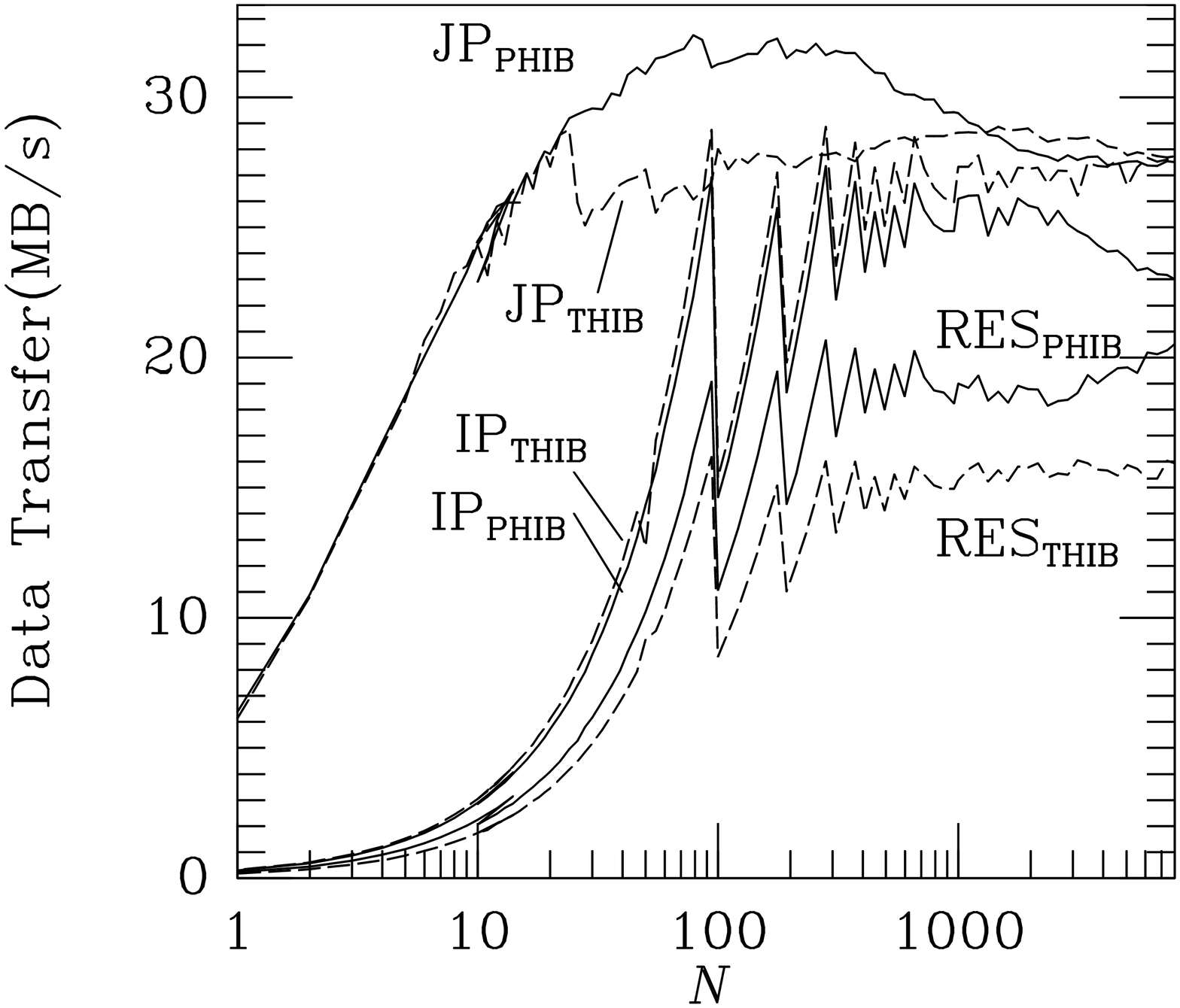}

\caption
{
The measured data transfer speed of library functions plotted as a
function of the number of particles.  The curves labeled IP, JP, and
RES indicate the performance of $i$-particles write operation,
$j$-particles write operation, and calculation result read operation,
respectively.  The performance of the PHIB system are shown in solid
curves, and that of the THIB system are shown in dashed curves.
}
\end{figure}

As described in section 2.2, the main body of data transfer in actual
simulation is divided into three parts: $j$-particle write operation,
$i$-particle write operation, and result read operation. Figure 6 shows
the measured speed of data transfer for these three operations as a
function of the number of particles. The host computer we used was
a DEC AlphaStation 5-500/500 with 128MB memory (AS500).

The curves for $j$-particles write operation is smooth for both of
PHIB and THIB, while that for $i$-particles write and result read
operations show discontinuous drop at $N = n_{\rm vp}$, $2 n_{\rm vp}
$, $3 n_{\rm vp} $..., where $n_{\rm vp}(=94)$ is the number of
virtual pipelines on PB. This is because the data transfer in those
operations are performed on block of $n_{\rm vp}$ particles. For $N$
which is not an exact integer multiple of $n_{\rm vp}$, part of the
data transferred in $i$-particle write and result read operations are
not used. Thus for $N \ll n_{\rm vp}$, the speed of these two
operations can be very low. To be more precise, the time to send $n$
$j$-particles can be approximated as
\begin{equation}
	\label{jp}
	T_{\rm j} = C_{\rm 1} + C_{\rm 2} n,
\end{equation}
where $C_{\rm 1} = $ 8.5 $\times$ 10${}^{-6}$ sec and $C_{\rm 2} = $
2.5 $\times$ 10${}^{-6}$ sec while that for $i$-particle
write and result read are
\begin{eqnarray}
	\label{ip}
	T_{\rm i} & = & C_{\rm i} 
	\left[ \frac{n + n_{\rm vp} - 1}{n_{\rm vp}} \right],
\end{eqnarray}
and
\begin{eqnarray}
	\label{rp}
	T_{\rm r} & = & C_{\rm r} 
	\left[ \frac{n + n_{\rm vp} - 1}{n_{\rm vp}} \right],
\end{eqnarray}
where $C_{\rm i} = $ 1.4 $\times$ 10${}^{-4}$ sec and $C_{\rm r} = $
1.9 $\times$ 10${}^{-4}$ sec. Here, $[x]$ denotes the maximum integer
which does not exceed $x$.

The peak performance of the data transfer through library functions is
somewhat lower than that of the raw data transfer.
For the read operation, this low performance is mainly due to the
cost of data transfer within the main memory of the host. After the
data read from PHIB are stored in a continuous memory block, each data
must be moved to appropriate address given by the user of the library.
To see the cost of data transfer within the main memory, we measured
the performance of the read operation without memory copy and got the
speed of $\sim$ 25MB/s.  This result can explain the difference
between the performance of the raw data read and that of the result
read operation.
For the write operations, the major reason for the performance
reduction is that the host has to fetch data from the main memory.  In
actual simulation, these data are scattered in large region of the
main memory ($\geq$ 10MB). In addition, the structure of the data to be
sent to PHIB and that of the data given by the user are
different. Thus the write operations have to fetch data from wide and
non-contiguous range of the main memory.
To see the cost of data fetch from main memory, we measured the write
speed in two different ways. We used data which occupy more than 50MB
of the main memory. In the measurement described in section 4.1, we
used data in the register of CPU.  In the first test, we sent them to
PHIB in contiguous order. In this measurement, the write speed was
reduced to $\sim$ 50MB/s. In the second, we used the same data but
sent them in non-contiguous order. In this case, the write speed was
reduced to $\sim$ 30MB/s. These results are consistent with the
performance of the write operations.

Compared to the transfer speed of these operations, the speed of Hlink
is high enough (64MB/s) and does not limit the performance of the PHIB
system.

\subsection{Direct Summation Code, Equal-timestep}

We performed test runs with an equal-timestep code. As the initial
conditions, we used a Plummer model.  We changed the number of
particles $N$ from 1024 to 262144 and measured the speed of
calculation with one PB.  The system of units is chosen so that the
total mass of the system $M$ and the gravitational constant $G$ are
both unity. The total energy of the system $E$ is $-1/4$ (Heggie and
Mathieu 1986).  The mass of all particles are $m=1/N$. The softening
parameter is $N^{-1/3}$, where $N$ is the number of particles. The
host computer was an AS500. We performed the same run with the THIB
and Alpha AXP 3000/700 (225MHz 21064A processor with 64MB memory,
hereafter AA3000).

\clearpage
\begin{figure}

\plotone{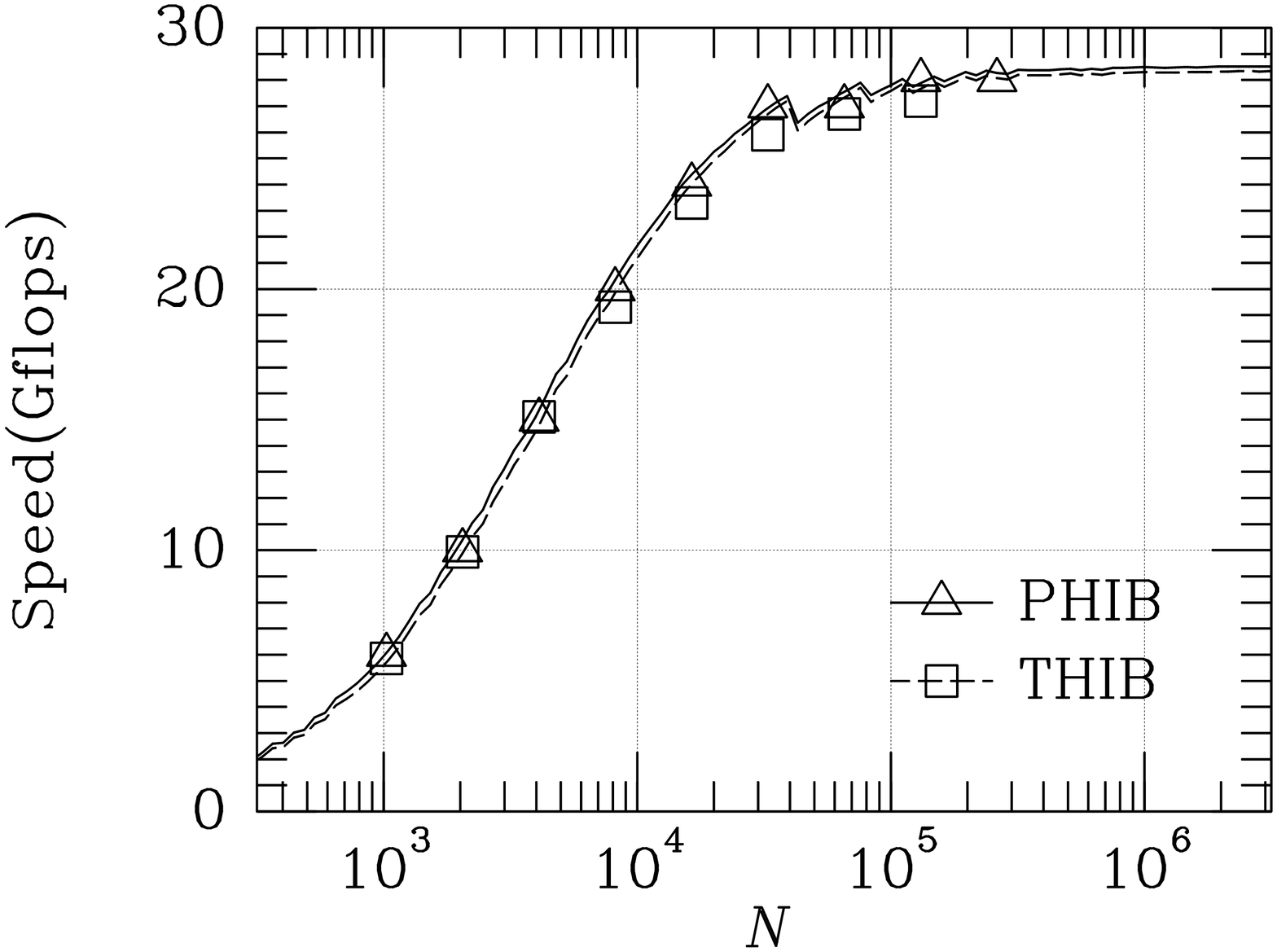}

\plotone{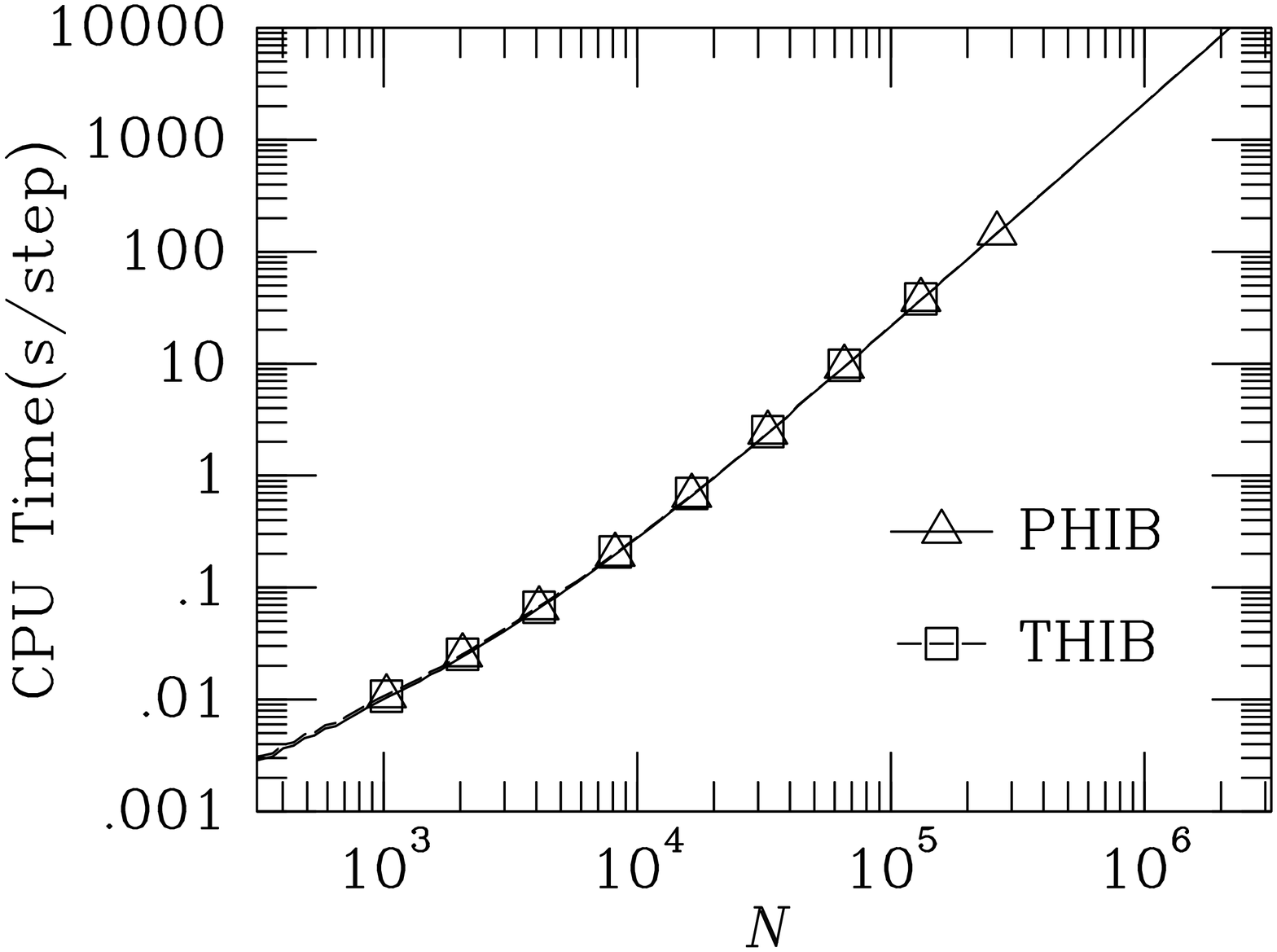}
\end{figure}

\begin{figure}
\plotone{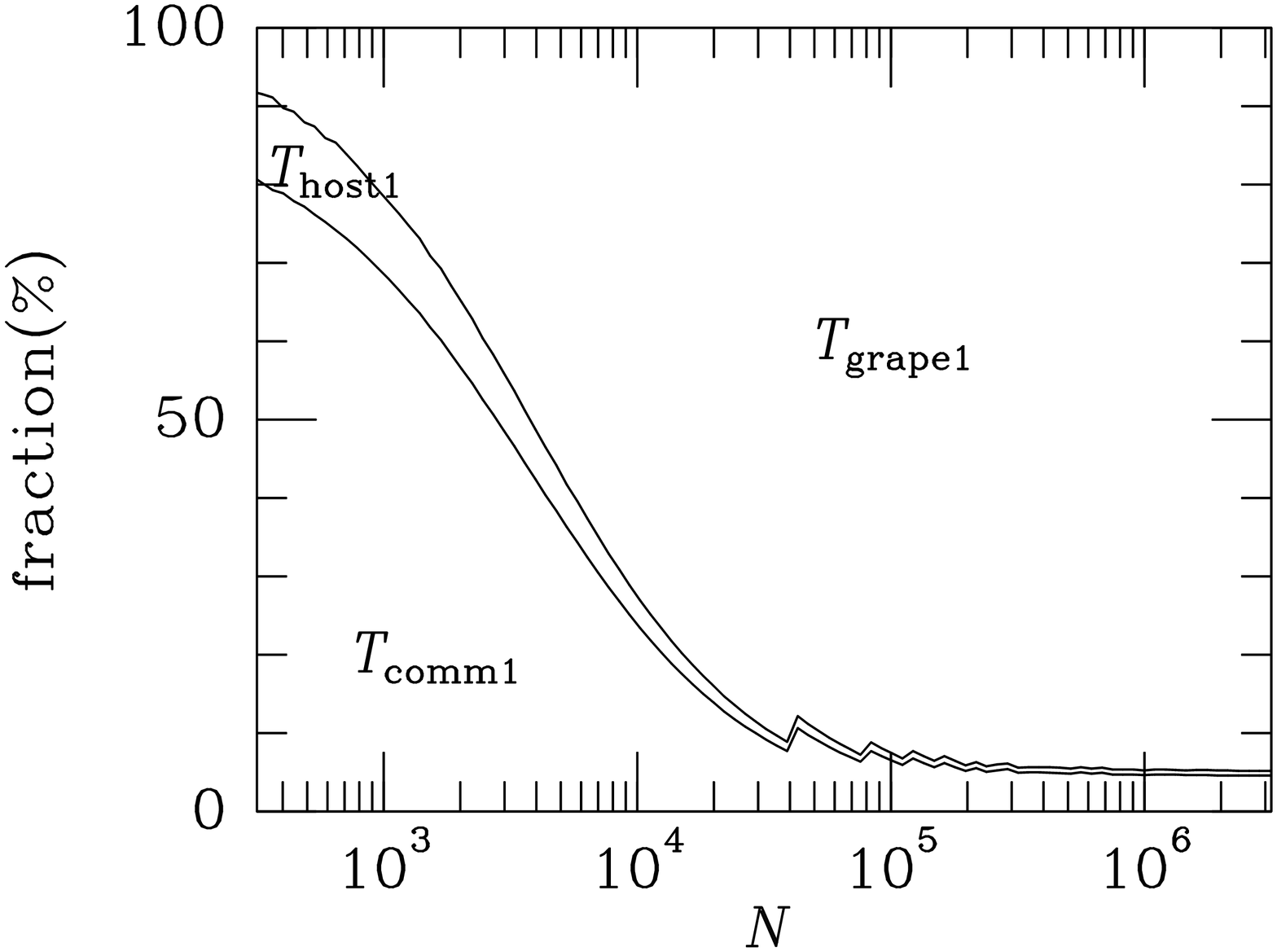}

\plotone{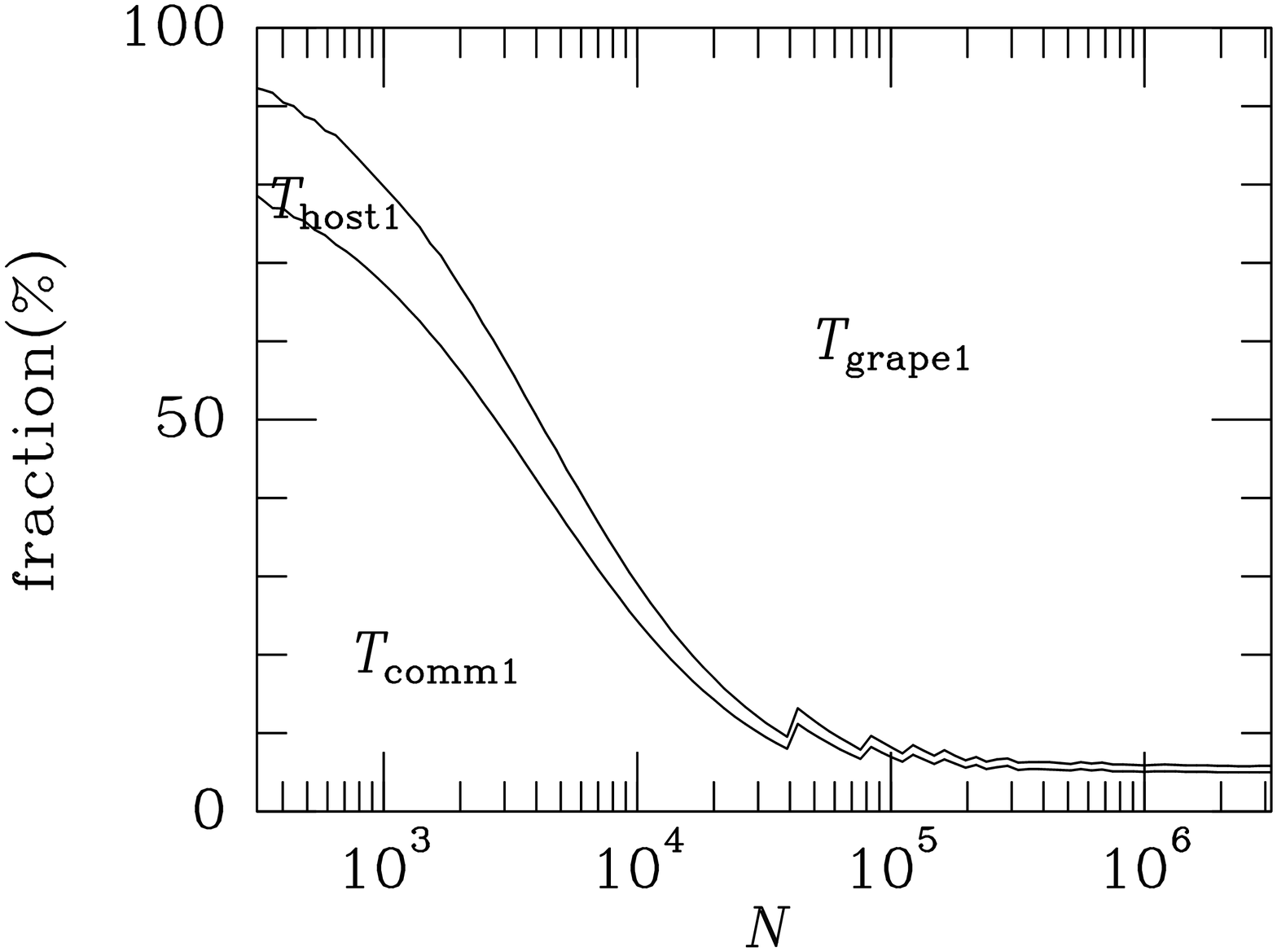}

\caption[ge.eps,pe.eps,fe.eps,fet.eps]
{
(a) The calculation speed and (b) CPU time per one timestep of GRAPE-4
for equal-timestep algorithm, plotted as a function of the number of
particles in the system. Solid and dashed curves represent the
theoretical estimate for PHIB and THIB, respectively. Triangles and
Squares represent the measured performance of PHIB and THIB,
respectively.  (c) The fraction of the time spent on the host
calculation, data transfer, and calculation on GRAPE for the PHIB
system, plotted as a function of the number of particles in the
system.  (d) Same as figure 7c, but for the THIB system.
}
\end{figure}

Figure 7a shows the calculation speed of GRAPE-4 in Gflops and figure 7b
shows the average CPU time per one timestep, for both PHIB and THIB.

The time to integrate the system of $N$ particles for one timestep is
estimated as follows:
\begin{equation}
	\label{equt}
	T_{\rm eq} = T_{\rm host1} + T_{\rm grape1} + T_{\rm comm1},
\end{equation}
where $T_{\rm host1}$, $T_{\rm grape1}$, and $T_{\em comm1}$ are the time spent on
host, the time spent on GRAPE-4, and the time spent for data transfer
between the host and GRAPE-4, respectively.  The three terms in the
right-hand side of equation (5) are expressed as
\begin{eqnarray}
	\label{equth}
	T_{\rm host1} & = & N t_{\rm host1},\\
	\label{equtg}
	T_{\rm grape1} & = & g_1 \frac{3 N^2 t_{\rm pipe}}{n_{\rm vp}},\\
	\label{equtc}
	T_{\rm comm1} & = & N (19 t_{\rm jp} + 10 g_1 r t_{\rm ip} +
	10 g_1 r t_{\rm res}).
\end{eqnarray}
Here, $t_{\rm host1}$ in equation (6) is the time for the host
computer to integrate one particle for one timestep.  In equation (7),
$n_{\rm vp}$ is the number of virtual pipelines on PB, and $t_{\rm
pipe}$ is the cycle time of PB. For AS500, $t_{\rm host1} = $ 0.5
$\times$ 10${}^{-6}$ sec and for AA3000, $t_{\rm host1} = $ 1.0
$\times$ 10${}^{-6}$ sec.  For the system we used, $n_{\rm
vp}=94$. The clock cycle of PB is 16 MHz, which corresponds to $t_{\rm
pipe}$ = 6.25 $\times$ 10${}^{-8}$ sec. The parameter $g_1$ is the
loss of parallel efficiency of multiple pipelines, which is estimated
as
\begin{equation}
	\label{g1}
	g_1 = \left[ \frac{N + n_{\rm vp} - 1}{n_{\rm vp}} \right]
	\frac{n_{\rm vp}}{N}.
\end{equation}
In equation (8), $t_{\rm jp}$, $t_{\rm ip}$, and $t_{\rm res}$ are the
time to transfer one word (4 bytes) in $j$-particle write operation,
$i$-particle write operation, and result read operation,
respectively. For these parameters, we use the values measured at $N =
10^4$ (See figure 6). When $N$ is larger than 43690, which is the size
of the $j$-particle memory of PB, we have to divide $j$-particles to
groups with the number of particles not exceeding 43690. We can
calculate the total forces by summing up the forces from each
groups.The parameter $r$ denotes the number of such groups, which is
given by
\begin{equation}
	\label{r}
	r = \left[ \frac{N - 1}{43690} + 1 \right].
\end{equation}

Figure 7c and 7d shows the theoretical estimate of fraction of the time
spent on the host calculation, data transfer, and calculation on GRAPE
for PHIB and THIB, respectively.

In figures 7a and 7b, the calculation time of the PHIB system is only
5\% faster than that of the THIB system.  This is because the host
calculation time in this algorithm is small ($\sim$10\%). The
communication performance with PHIB is almost same as that with THIB
and the difference of these two system are mainly exist in the
performance of the hosts. Since the calculation time spent on the host
is small, the performance of the two systems are not so much
different.

\subsection{Direct Summation Code, Individual-timestep}

We performed test runs with an individual-timestep code. The initial
condition and hardware configuration are the same as the ones used in
the test run of equal-timestep code described in section 4.3, except
that the softening parameter is $2/N$. We changed the number of
particles from 1024 to 32768, and measured the speed of calculation.

\begin{figure}

\plotone{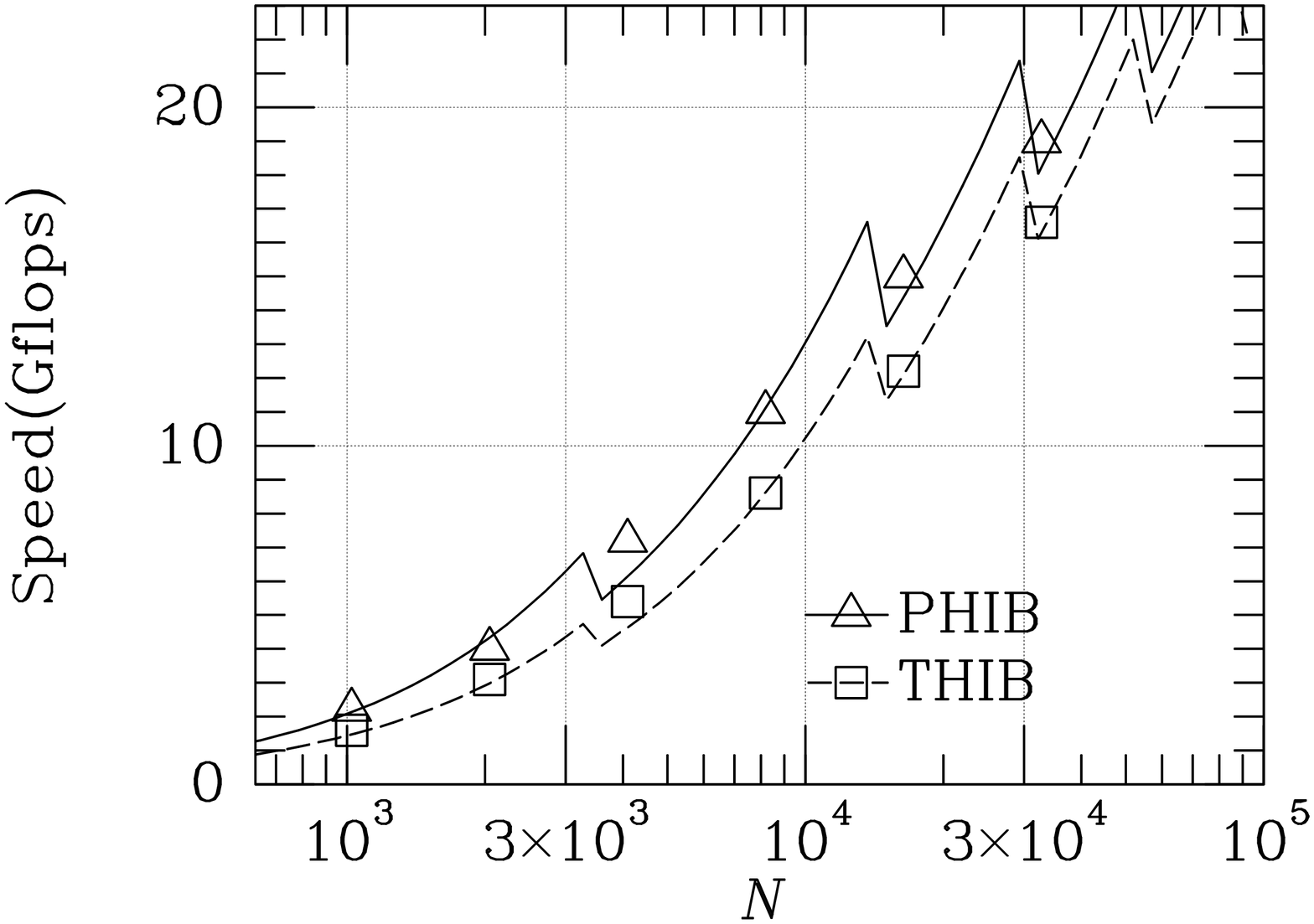}

\plotone{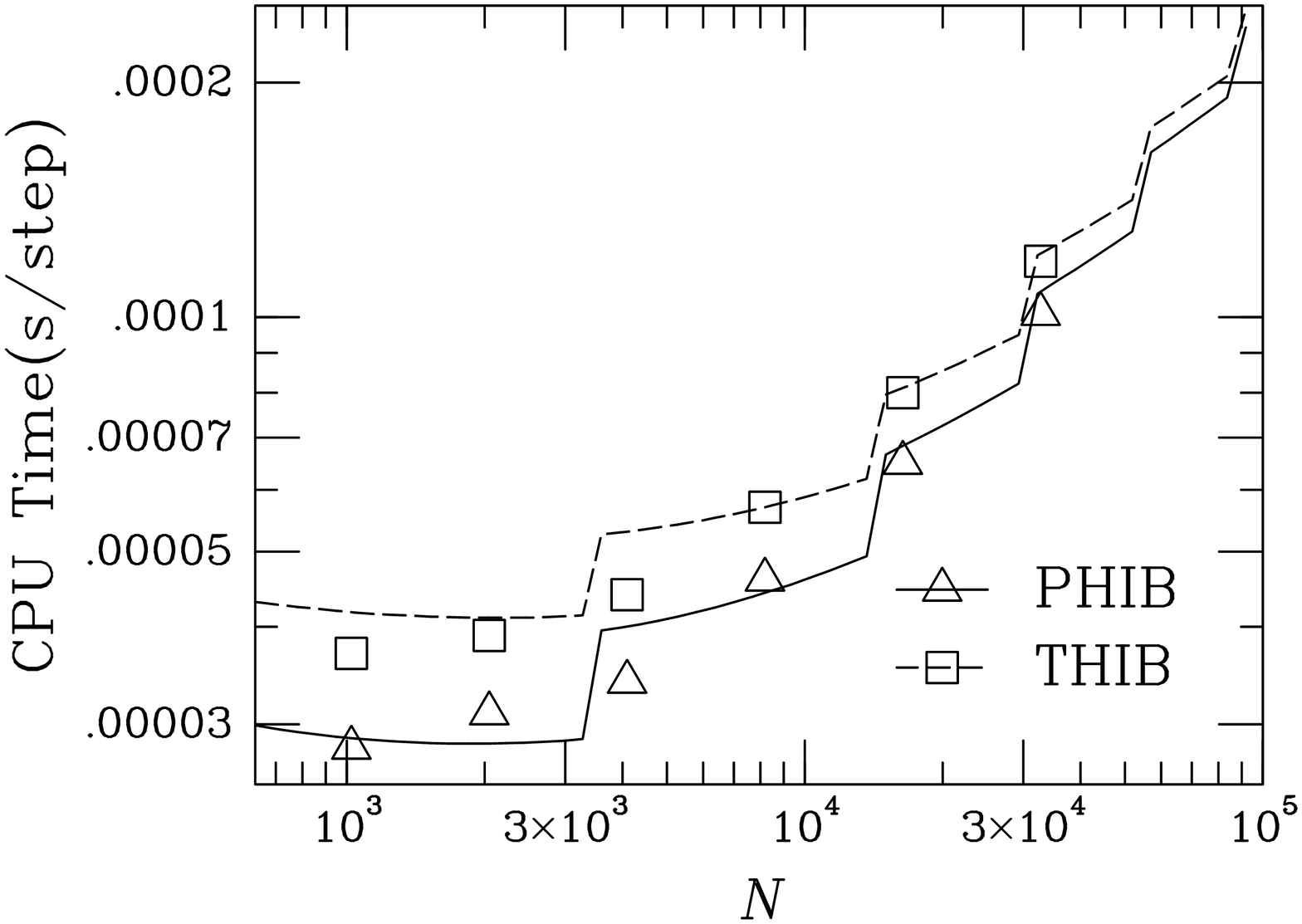}

\plotone{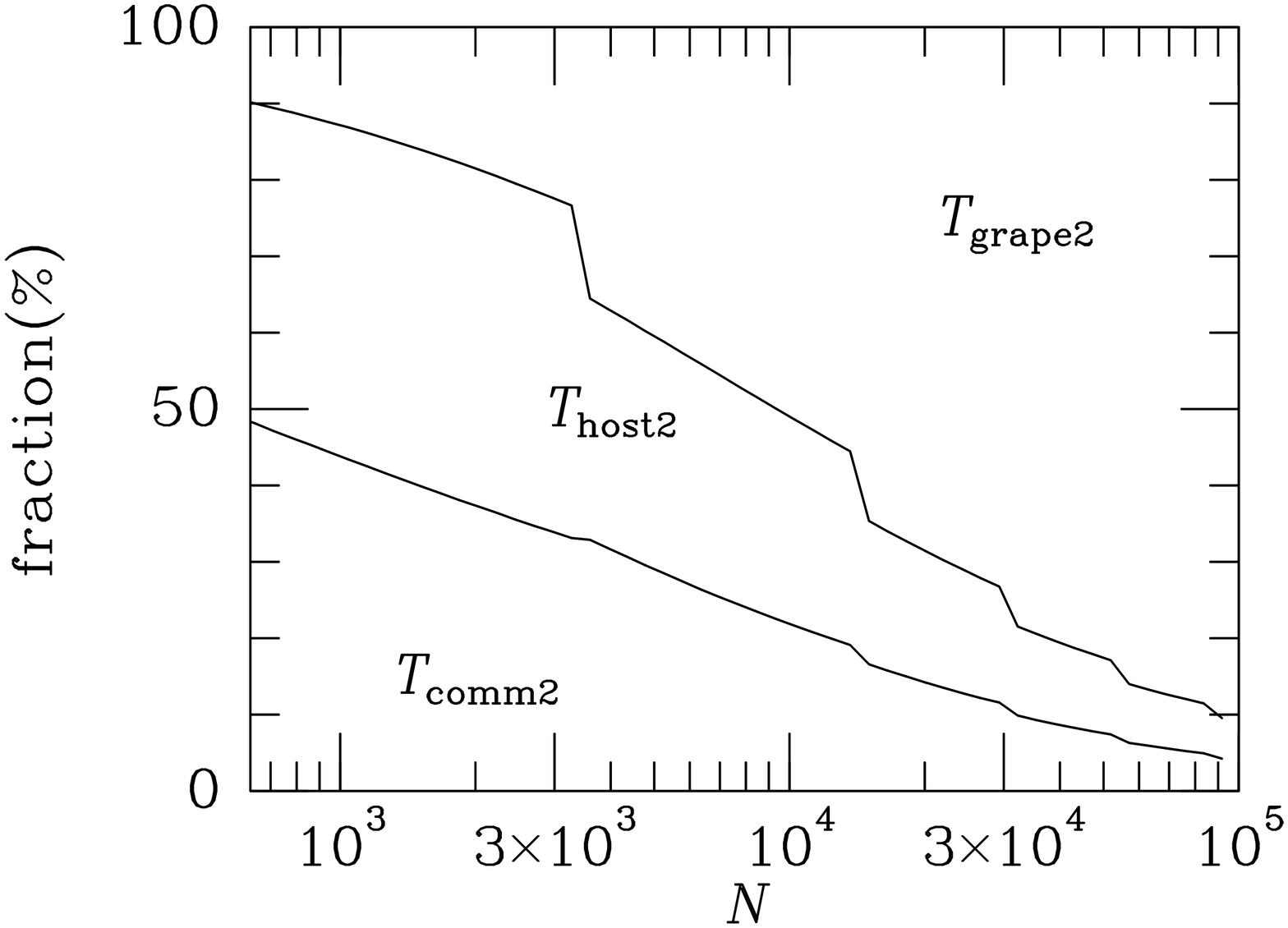}

\plotone{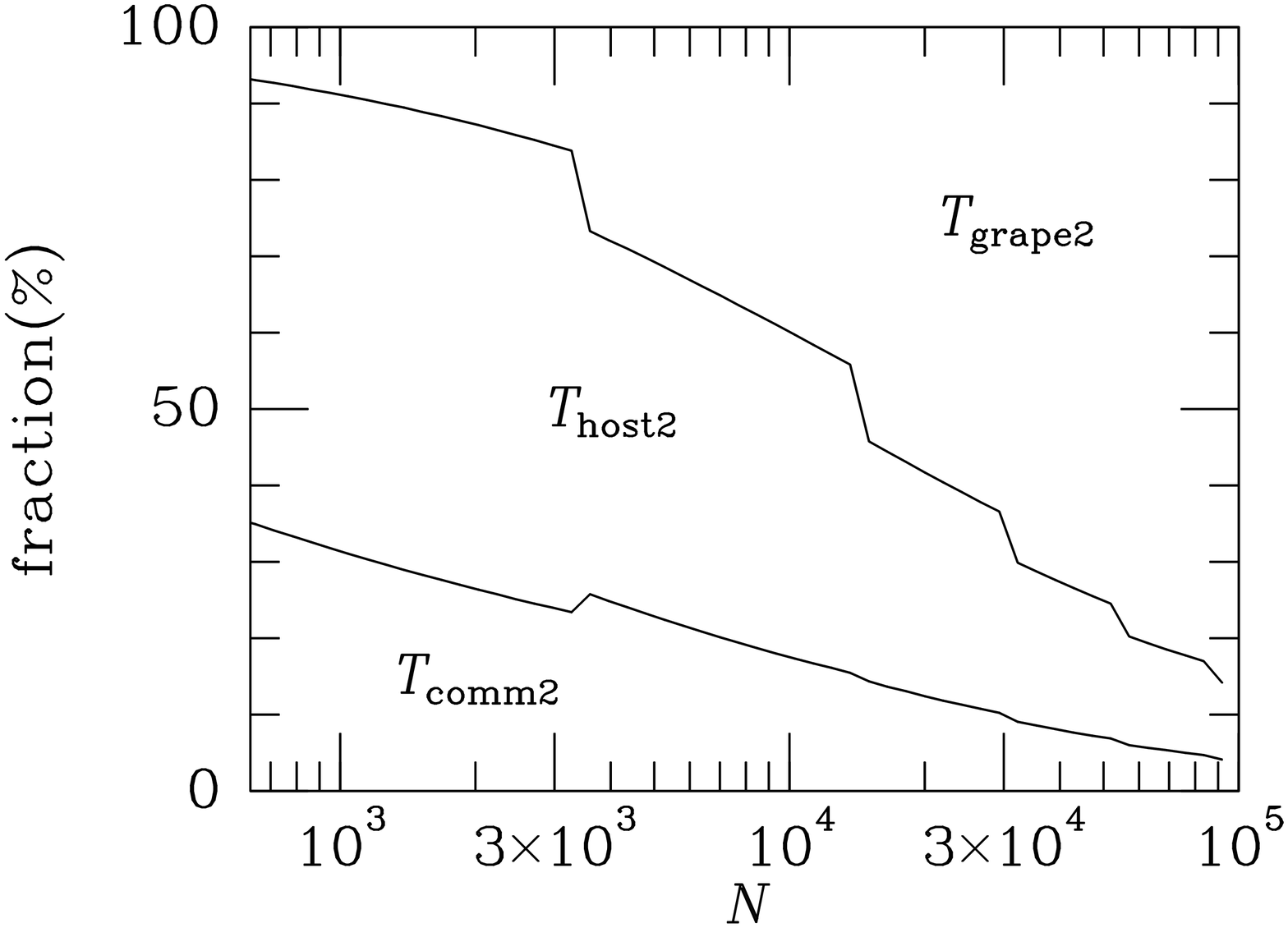}

\caption
{
Same as figure 7 but for the individual-timestep algorithm.
}
\end{figure}

Figure 8a shows the calculation speed of GRAPE-4 in Gflops and figure
8b shows the average CPU time to integrate one particle for one
timestep, for both PHIB and THIB.

The time to integrate one particle for one timestep is estimated as
follows:
\begin{equation}
	\label{idtt}
	T_{\rm idt} = T_{\rm host2} + T_{\rm grape2} + T_{\rm comm2},
\end{equation}
where $T_{\rm host2}$, $T_{\rm grape2}$, and $T_{\rm comm2}$ are the time spent on
host, the time spent on GRAPE-4, and the time spent for data transfer
between the host and GRAPE-4, respectively.  In equation
(11), the term $T_{\rm host2}$ is estimated as 1.3 $\times$
10${}^{-5}$ sec for AS500 and 2.5 $\times$
10${}^{-5}$ sec for AA3000.

The terms $T_{\rm grape2}$ and $T_{\rm comm2}$ are expressed as
\begin{eqnarray}
	\label{idttg}
	T_{\rm grape2} & = & g_2 \frac{3 N t_{\rm pipe}}{n_{\rm vp}},\\
	\label{idttc}
	T_{\rm comm2} & = & (38 t_{\rm jp} + 10 g_2 t_{\rm ip} + 10 g_2 t_{\rm res}).
\end{eqnarray}
Here, the parameter $g_2$ in equation (12) is the loss of
parallel efficiency of multiple pipelines, which is estimated as
\begin{equation}
	\label{g2}
	g_2 = \left[ \frac{n_{\rm s} + n_{\rm vp} - 1}{n_{\rm vp}} \right]
	\frac{n_{\rm vp}}{n_{\rm s}}.
\end{equation}
The parameter $n_{\rm s}$ is the average number of particles to share the
same time. For this parameter, we used experimental result (Makino et
al. 1997)
\begin{equation}
	\label{ns}
	n_{\rm s} = 1.6 N^{1/2}.
\end{equation}

In the case of the individual-timestep algorithm, the PHIB system is
10-30\% faster than the THIB system.  This is because the the host
calculation term $T_{\rm host2}$ in equation (11) are comparable to
$T_{\rm grape2}$ and $T_{\rm comm2}$ (See figure 8c and 8d). The
reduction of the host calculation time with this algorithm is more
significant than that with the equal-timestep algorithm.

\subsection{Barnes-Hut Treecode}

We performed test runs with a Barnes-Hut treecode modified for GRAPE
(Makino 1991a). The initial condition and the hardware configuration
are the same as the ones used in the test run of equal-timestep code
described in section 4.3. The opening angle for cells of treecode,
$\theta$, is 0.75, and the critical particle number for Barnes'
vectorization scheme, $n_{\rm crit}$, is 4000. Strictly speaking,
$n_{\rm crit}$ depends on $N$ and must be optimized for each value
of $N$ (Barnes 1990). However, the dependency of $n_{\rm crit}$ on $N$ is
weak, and we found that $n_{\rm crit} = $ 4000 is close to optimum for the
entire range of $N$ we used.  We changed the number of particles from
16384 to 1048576, and measured the speed of calculation.

\begin{figure}

\plotone{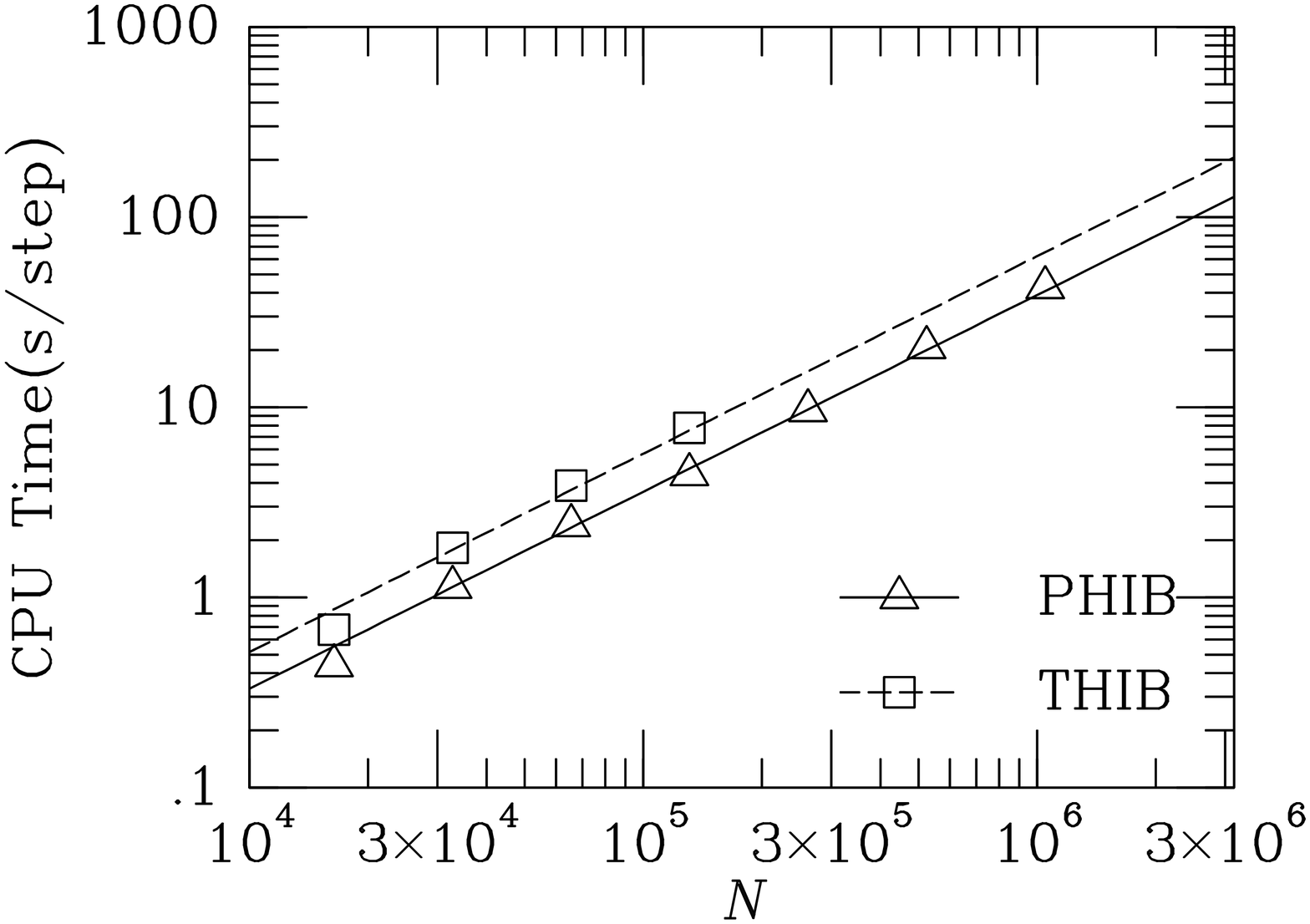}

\plotone{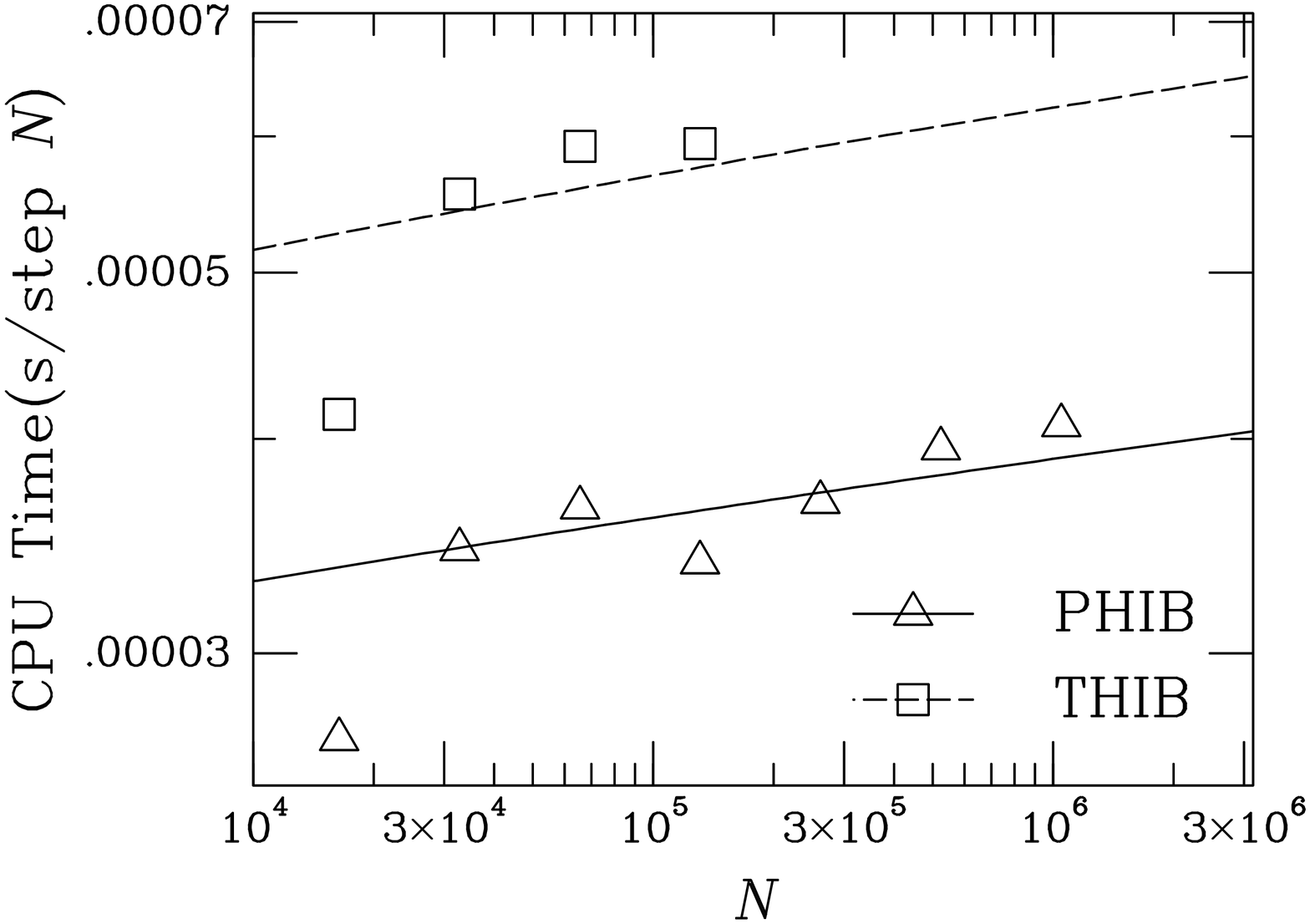}

\plotone{ftftt.eps}

\caption
{
(a)Same as figure 7b, but for the Barnes-Hut treecode.  (b)CPU time
per one timestep per one particle for the Barnes-Hut treecode. The
meaning of curves and symbols are the same as those in figure 9a.
(c)Theoretical estimate of CPU time per one timestep spent on GRAPE,
the host calculation, and data transfer for both PHIB and THIB.  The
number of particles is 1048576.
}
\end{figure}

Figure 9a shows the average CPU time per one timestep, for both PHIB
and THIB. Figure 9b shows the CPU time per step per particle. On
average, the PHIB system is about 40\% faster than the THIB system.

The calculation time for one timestep is estimated as follows:
\begin{equation}
	\label{treet}
	T_{\rm tree} = T_{\rm host3} + T_{\rm grape3} + T_{\rm comm3},
\end{equation}
where $T_{\rm host3}$, $T_{\rm grape3}$, and $T_{\rm comm3}$ are the time spent on
host, the time spent on GRAPE-4, and the time spent for data transfer
between the host and GRAPE-4, respectively. These terms are expressed
as
\begin{eqnarray}
	\label{treeth}
	T_{\rm host3} & = & N t_{\rm int} + 
		(N \log_{10} N) t_{\rm const} \nonumber \\
& &		+ \frac{N n_{\rm terms}}{n_{\rm g}} t_{\rm list},\\
	\label{treetg}
	T_{\rm grape3} & = & g_3 \frac{3 N n_{\rm terms} t_{\rm pipe}}{n_{\rm vp}},\\
	\label{treetc}
	T_{\rm comm3} & = & N (19 \frac{n_{\rm terms}}{n_{\rm g}}t_{\rm jp} +
    \nonumber \\
	& & 10 g_3 t_{\rm ip} + 10 g_3 t_{\rm res}).
\end{eqnarray}
In equation (17), $t_{\rm int}$ is the time for the host computer to
integrate one particle for one timestep, $t_{\rm const}$ is the time
to construct the tree structure, and $t_{\rm list}$ is the time to
create the interaction lists (Fukushige et al. 1991). These values are
estimated to be 1.9 $\times$ 10${}^{-6}$ sec, 1.6 $\times$ 10${}^{-6}$
sec, and 0.8 $\times$ 10${}^{-6}$ sec for AS500 and 3.6 $\times$
10${}^{-6}$ sec, 3.5 $\times$ 10${}^{-6}$ sec, and 3.3 $\times$
10${}^{-6}$ sec for AA3000, respectively. According to Makino(1991a),
the average length of the interaction list, $n_{\rm terms}$, and the
number of particles in the group, $n_{\rm g}$, are estimated as
\begin{eqnarray}
	\label{ng}
	n_{\rm g} & \simeq & n_{\rm crit} / 4,\\
	\label{nt}
n_{\rm terms} & \simeq & n_{\rm g} +
		14n_{\rm g}^{2/3} +
		84n_{\rm g}^{1/3} +
		56\log_8 n_{\rm g}  \nonumber\\
& &		 - 31 \theta^{-3} \log_{10}n_{\rm g} - 72  \nonumber\\
& &		+ 10^2 \theta^{-3} \log_8 10{\frac{N\theta^3}{23}}.
\end{eqnarray}
In equation (18), $g_3$ is the loss of parallel efficiency
of multiple pipelines, which is estimated as
\begin{equation}
	\label{g3}
	g_3 = \left[ \frac{n_{\rm g} + n_{\rm vp} - 1}{n_{\rm vp}} \right]
	\frac{n_{\rm vp}}{n_{\rm g}}.
\end{equation}

\section{Summary and Discussion}

In this paper, we presented the design and performance of PHIB, the
PCI interface for the GRAPE-4 special-purpose computer. It allows us
to use a wide variety of host computers, from Intel-based PCs to vector
supercomputers.

\begin{figure}

\plotone{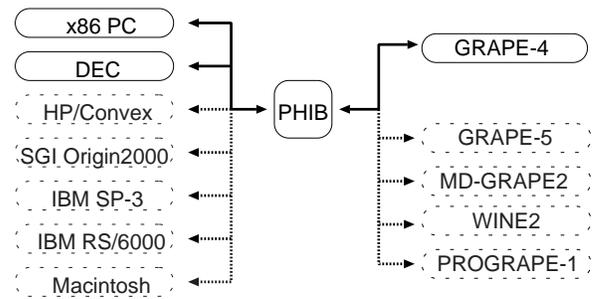}

\caption
{
Future plan for PHIB. PHIB can be used to connect various host
computers to GRAPE systems to be developed in future.
}
\end{figure}

Another important implication of PHIB is that we can use it as a
generic host interface for our future projects. The protocol used in
Hlink is quite simple and does not require sophisticated
implementation, either in logical or electrical designs, and yet it
can achieve the data transfer speed up to 100 MB/s, with extremely
small latency (the latency of a PIO read operation is less than 500
ns). We plan to use PHIB as the interface to new systems we will
develop in the next few years (See figure 10). \par
\vspace{1pc}\par

\acknowledgments

We thank Evangelie Athanassoulas for providing the host computer of
GRAPE-4 system in performance measuring. A.K. would like to thank
Evangelie Athanassoulas for her kind hospitality during his stay at
Marseilles. This work was supported by Grant-in-Aid for Specially
Promoted Reserch (04102002) of the ministry of Education, Science,
Spotrts and Culture, Japan, and by National Science Foundation under
an Advanced Scientific Computing grant ASC-9612029.

\end{document}